# Transmission Line Inspires A New Distributed Algorithm to Solve the Nonlinear Dynamical System of Physical Circuit

Fei Wei

*weifei00@mails.tinghua.edu.cn*

Huazhong Yang

*yanghz@tsinghua.edu.cn*

*Abstract*

As known, physical circuits, e.g. integrated circuits or power system, work in a distributed manner, but these circuits could not be easily simulated in a distributed way. This is mainly because that the dynamical system of physical circuits is nonlinear and the linearized system of physical circuits is nonsymmetrical. This paper proposes a simple and natural strategy to mimic the distributed behavior of the physical circuit by mimicking the distributed behavior of the internal wires inside this circuit.

Mimic Transmission Method (MTM) is a new distributed algorithm to solve the nonlinear ordinary differential equations extracted from physical circuits. It maps the transmission delay of interconnects between subcircuits to the communication delay of digital data link between processors.

MTM is a black-box algorithm. By mimicking the transmission lines, MTM seals the nonlinear dynamical system within the subcircuit. As the result, we do not need to pay attention on how to solve the nonlinear dynamic system or nonsymmetrical linear system in parallel.

MTM is a global direct algorithm, and it does only one distributed computation at each time window to obtain accurate result, so unconvergence issues do not need to be worried about.

*Key Words:*

Parallel Computing, Nonlinear Ordinary Differential Equations, Delay Differential Equations, Wave Equation, Distributed Simulation, Integrated Circuit, Power Supply Network, Transmission Line, Wire, Interconnect, Wire Tearing, Parallel SPICE, Transistor-level Full-chip Simulation

# 1. Introduction

Since 1980s, the distributed simulation of integrated circuits became a hot topic [3][4][5][6][7][8][9]. Recently, many start-ups dedicated themselves into this challenging work. The mathematic description of physical circuits, e.g. integrated circuit or power system, is a large set of nonlinear ordinary differential equations (ODE), and SPICE is an excellent nonlinear solver for this kind of problems [1][2]. Fig. 1 shows the work flow of the transient analysis in SPICE.

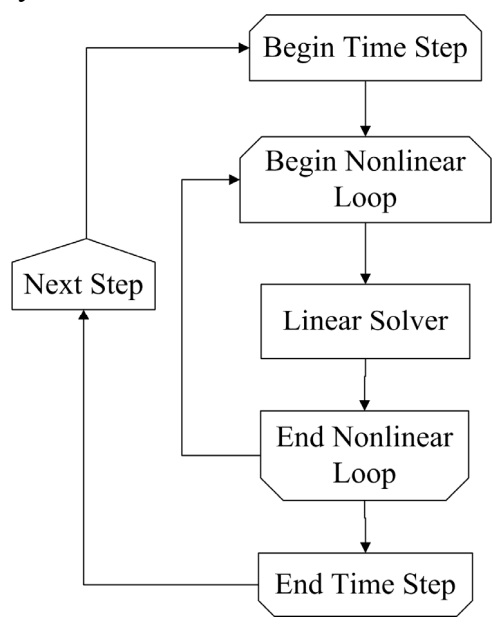

Figure 1. Work flow of the transient analysis in SPICE.

To simulate the circuit in parallel, one strategy is the Waveform Relaxation method (WR) [3][4]. WR is impractical because its convergence speed is too slow. Nowadays the most prevalent strategy is the distributed Newton-Raphson method (NR) [5][6][7][8][9][10][11]. The procedure of the distributed NR method is illustrated in Fig. 2 [8]. First it discretizes the dynamical system into a nonlinear system; then it linearizes this nonlinear system into a linear system; finally it solves this linear system in parallel. The shortage of the distributed NR method is that frequent distributed iterations make this algorithm inefficient [7].

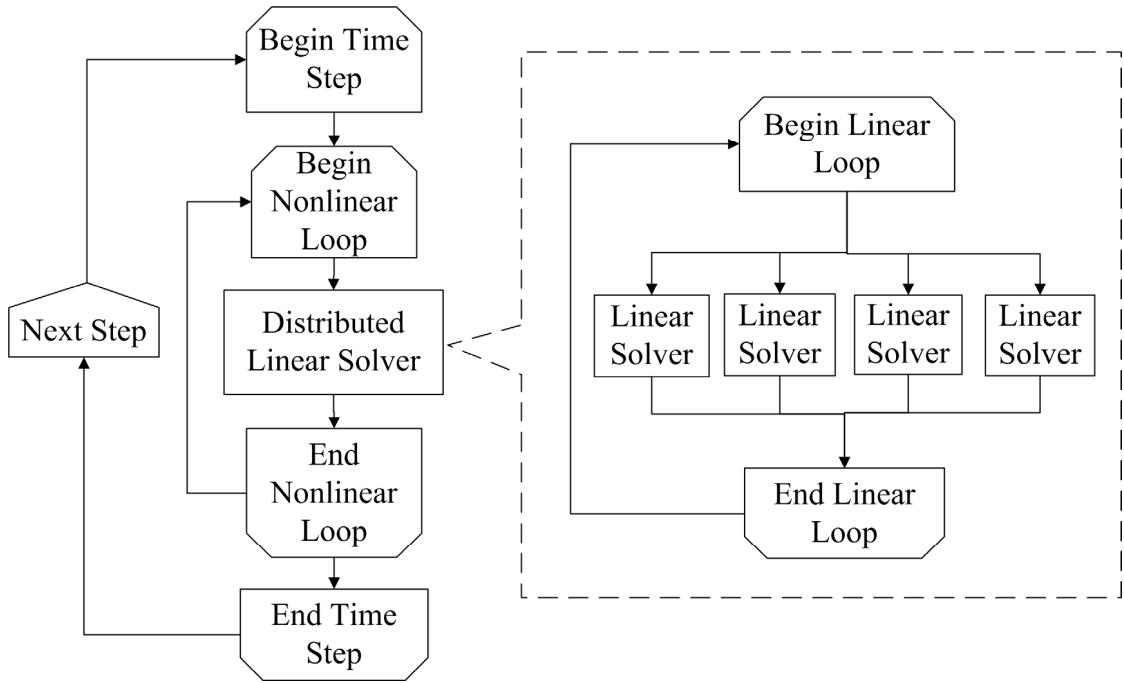

Figure 2. Work flow of the distributed NR method.

For most cases, the linearized system of physical circuits is nonsymmetrical, because of the existence of the controlled sources in the circuit [8]. Nonsymmetrical linear system is not easy to be solved in parallel, and Schur complement method is frequently used [7][10][11]. This method makes use of the master-slave model, and thus its scalability is limited [23].

Many efforts have been made by us for the distributed computing of linear or dynamical system extracted from circuits [25][26][27][28]. An important observation is that the transmission line (or wire, interconnects) plays a key role for the scalability and stability of the distributed physical circuits.

Virtual Transmission Method (VTM) is an efficient and scalable distributed algorithm to solve the sparse linear system of resistor networks on arbitrary number of processors [25][28]. Waveform Transmission Method (WTM) is a waveform relaxation based algorithm to solve ordinary differential equations of resistor-capacitor network in parallel [27].

The shortage of VTM is that, when solving the nonsymmetrical linear system, this algorithm might be out of convergence, if the character impedances of the virtual transmission lines are not proper selected. This shortage limits the application of VTM to simulate integrated circuits and power systems.

Recently, we come to realize that, it is not necessary to artificially add virtual transmission lines into the system, because transmission lines (or wires, interconnects) are inherent and everywhere in the physical circuits. As the result, we might use the internal wires inside integrated circuits to partition the system and isolate different subcircuits.

In addition, if we do not neglect the tiny transmission delay of the wires inside the circuit, the mathematic description of physical circuit would actually be nonlinear delay differential equations (DDE), which consist of the nonlinear ordinary differential equations and wave equations.

Mimic Transmission Method (MTM) is a new distributed numerical algorithm to solve nonlinear dynamical system of physical circuits. As a distributed algorithm, MTM totally mimics the distributed behavior of the physical circuit by mimicking the distributed behavior of the internal wires inside this circuit. Fig. 3 shows the work flow of MTM. MTM is a black box algorithm and we do not need to know the details on how to solve the nonlinear ODE inside this black box.

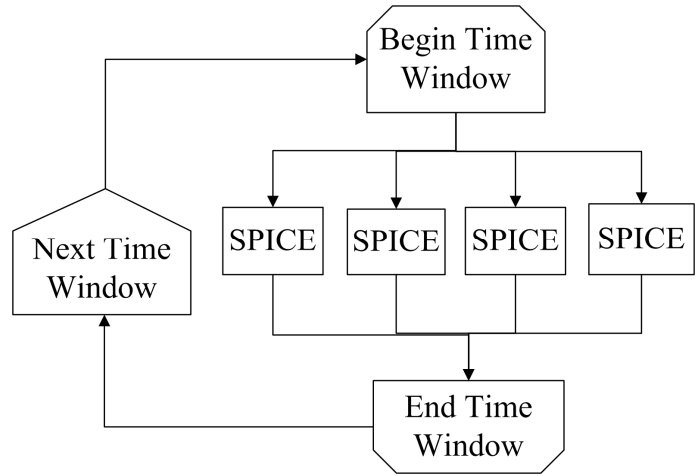

Figure 3. Work flow of Mimic Transmission Method.

The basic idea is to partition the physical circuit by the internal wires inside this circuit. Then, these internal wires are mimicked by the digital data link between processors. In this case, we do not need to pay much attention on how to optimize the characteristic impedances of these wires (as what we did in VTM), and we just set the characteristic impedances as the same as the value extracted from the physical circuit.

In this paper, we classify distributed numerical algorithms into two categories: global iterative algorithm and global direct algorithm. Theoretically, if the algorithm could obtain the exact answer within one or a limited number of distributed computations, it is a global direct algorithm, e.g. Schur complement method, ScaLAPACK; if the algorithm should perform unlimited number of distributed computations to approach the exact answer, it is a global iterative algorithm, e.g. Block-Jacobi, VTM, WR, WTM. Consequently, with the background of parallel computing, sequential algorithms running on a single processor is called local algorithms. If we do the Newton-Raphson iterations (NR) on a single processor, it is a local iterative algorithm; if we do the Newton-Raphson iterations (NR) on a number of processors, it is a global iterative algorithm. Fig. 4 shows this classification for numerical algorithms.

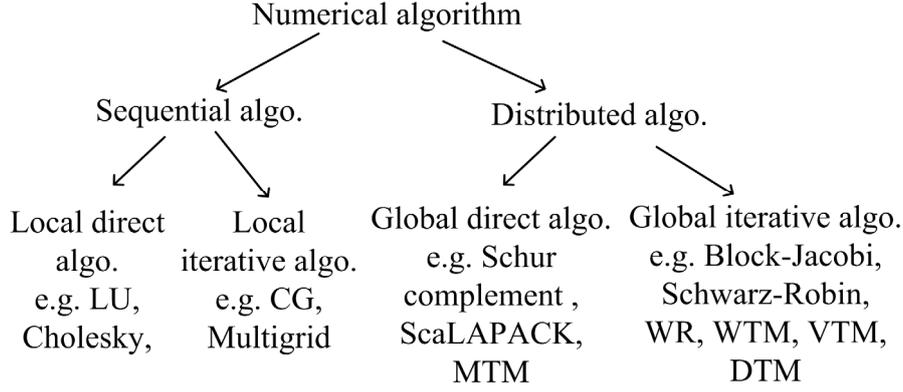

Figure 4. Classification of numerical algorithms.

This paper is organized as follows. Section 2 gives the basics of transmission line. Section 3 describes the basic idea of delay mapping. Section 4 gives a simple example to show the procedure of MTM. Section 5 makes some comments. We conclude this work in Section 6.

## 2. Transmission Line

In this paper, transmission line, wire and interconnect have the same meaning. Transmission line is a linear electric element, and is ubiquitous in the physical circuit, i.e. integrated circuit, power supply network and microwave / radio-frequency circuit. It could be classified into lossless lines and lossy lines.

## 2.1 Lossless Transmission Line

The circuit diagram of the transmission line is illustrated in Fig. 5. The mathematical description of the lossless or ideal transmission line is the telegrapher equations:

$$\frac{\partial u}{\partial x} = -L\frac{\partial i}{\partial t} \tag{2.1}$$

$$\frac{\partial i}{\partial x} = -C\frac{\partial u}{\partial t} \tag{2.2}$$

Here $L$ is the inductance per unit length, $C$ is the capacitance per unit length.

By taking $\frac{\partial}{\partial x}$ of (2.1) and take $\frac{\partial}{\partial t}$ of (2.2), we get the wave equations:

$$\frac{\partial^2 u(x,t)}{\partial^2 x} = LC\frac{\partial^2 u(x,t)}{\partial^2 t} \tag{2.3}$$

Also, by taking $\frac{\partial}{\partial t}$ of (2.1) and take $\frac{\partial}{\partial x}$ of (2.2), we get (2.4).

$$\frac{\partial^2 i(x,t)}{\partial^2 x} = LC\frac{\partial^2 i(x,t)}{\partial^2 t} \tag{2.4}$$

The transmission velocity of the voltage wave or current wave is:

$$v = 1/\sqrt{LC}$$

The time domain solution of the lossless transmission line is in (2.5), which is called Transmission Delay Equations [29][30][31][32]. The equivalent circuit of the lossless transmission line is illustrated in Fig. 5B.

$$u(0,t) + Z \cdot i(0,t) = u(l, t-\tau) + Z \cdot i(l, t-\tau)$$
$$u(l,t) - Z \cdot i(l,t) = u(0, t-\tau) - Z \cdot i(0, t-\tau) \quad (2.5)$$

Here $l$ is the length of the lossless transmission line. $u(x,t)$ and $i(x,t)$ are the voltage and current at point $x$ of the line at the time $t$, $x \in [0,l]$. $\tau$ is the propagation delay.

$$\tau = \frac{l}{v} = l\sqrt{LC} \quad (2.6)$$

$Z$ is the characteristic impedance,

$$Z = \sqrt{\frac{L}{C}} \quad (2.7)$$

If we set:

$$u_1(t) = u(0,t) \quad i_1(t) = i(0,t)$$
$$u_2(t) = u(l,t) \quad i_2(t) = -i(l,t)$$

Then (2.5) is re-expressed as (2.8).

$$u_1(t) + Z \cdot i_1(t) = u_2(t-\tau) - Z \cdot i_2(t-\tau)$$
$$u_2(t) + Z \cdot i_2(t) = u_1(t-\tau) - Z \cdot i_1(t-\tau) \quad (2.8)$$

Here $u_1(t)$ and $u_2(t)$ represent the ports' voltages, and $i_1(t)$ and $i_2(t)$ represent ports' inflow currents.

## 2. 2 Lossy Transmission Line

The mathematical description of the lossy transmission line is (2.9) and (2.10), according to the telegrapher equations, which is linear partial differential equations:

$$\frac{\partial u}{\partial x} = -L\frac{\partial i}{\partial t} - R \cdot i \quad (2.9)$$

$$\frac{\partial i}{\partial x} = -C\frac{\partial u}{\partial t} - G \cdot u \quad (2.10)$$

The boundary conditions are:

$$u(x,t) = 0, i(x,t) = 0, x \in [0,l], t \in (-\infty, 0]$$

Merge (2.9) and (2.10), we get the wave equations with linear dissipative terms:

$$\frac{\partial^2 u}{\partial x^2} = LC\frac{\partial^2 u}{\partial t^2} + (LG + RC)\frac{\partial u}{\partial t} + RG \cdot u$$

$$\frac{\partial^2 i}{\partial x^2} = LC\frac{\partial^2 i}{\partial t^2} + (LG + RC)\frac{\partial i}{\partial t} + RG \cdot i$$

The time domain solution for the lossy transmission line is much more complicated [20]. (2.11) is the transmission delay equations for the lossy transmission lines.

$$u_1(t) + Z \cdot i_1(t) + u_1(t) * h(t) = e^{-\beta\tau} \begin{Bmatrix} u_2(t-\tau) - Z \cdot i_2(t-\tau) + \\ u_2(t-\tau) * g(t) - Z \cdot i_2(t-\tau) * f(t) \end{Bmatrix}$$

$$u_2(t) + Z \cdot i_2(t) + u_2(t) * h(t) = e^{-\beta\tau} \begin{Bmatrix} u_1(t-\tau) - Z \cdot i_1(t-\tau) + \\ u_1(t-\tau) * g(t) - Z \cdot i_1(t-\tau) * f(t) \end{Bmatrix}$$
(2.11)

Here * is the convolution operator, for example, $x(t) * y(t) = \int_{-\infty}^{+\infty} x(s)y(t-s)ds$.

$h(t)$, $g(t)$ and $f(t)$ are known functions:

$$h(t) = e^{-\beta t}\alpha [I_1(\alpha t) - I_0(\alpha t)]$$

$$g(t) = e^{-\beta t}\alpha u(t) \left[ \frac{t+\tau}{\sqrt{t^2 + 2\tau t}} I_1(\alpha\sqrt{t^2 + 2\tau t}) - I_0(\alpha\sqrt{t^2 + 2\tau t}) \right]$$

$$f(t) = e^{-\beta t}\alpha u(t) \frac{\tau}{\sqrt{t^2 + 2\tau t}} I_1(\alpha\sqrt{t^2 + 2\tau t})$$

$$\beta = \frac{1}{2}(\frac{R}{L} + \frac{G}{C}), \quad \alpha = \frac{1}{2}(\frac{R}{L} - \frac{G}{C}),$$

$$\tau = l\sqrt{LC}.$$

Where $u(t)$ is the unit step function. $I_0(t)$ and $I_1(t)$ are the modified Bessel function of the zeroth and first order [33]. $I_0'(t) = I_1(t)$.

If we set $R = 0$, $G = 0$, then $\alpha = 0$, $\beta = 0$, and (2.11) is degenerated into (2.8). According to (2.11), we conclude that the signal propagation delay of the lossy transmission line is the same as the lossless line.

Then we consider the numerical computation of the convolution integral. Assume $x(t)$ is the unknown, and $y(t)$ is a known function. $x(t) = 0$, $y(t) = 0$, when $t \in (-\infty, 0)$. $\Delta t = \frac{t}{k}, k \in \mathbb{N}^+$.

$$x(t) * y(t) = \int_{-\infty}^{+\infty} x(s)y(t-s)ds = \int_{0}^{t} x(s)y(t-s)ds$$

$$= \sum_{i=0}^{k-1} \int_{i\Delta t}^{(i+1)\Delta t} x(s)y(t-s)ds$$

$$= \sum_{i=0}^{k-2} \int_{i\Delta t}^{(i+1)\Delta t} x(s)y(t-s)ds + \int_{t-\Delta t}^{t} x(s)y(t-s)ds$$

$$\approx \sum_{i=0}^{k-2} \int_{i\Delta t}^{(i+1)\Delta t} x(s)y(t-s)ds + x(t)y(0)\Delta t$$

$$= C_1 \cdot x(t) + C_2$$

Where the unknown at the time $t$ is $x(t)$. $C_2 = \sum_{i=0}^{k-2} \int_{i\Delta t}^{(i+1)\Delta t} x(s)y(t-s)ds$ is a known value and it is a linear combination of the old calculated values of $x(s), s \in [0,t)$. $C_1 = y(0)\Delta t$ is a constant [21].

Based on the conclusion above, we simplified (2.11) into (2.12), which is the discrete form of (2.11).

$$\begin{aligned} A_1(t) \cdot u_1(t) + B_1(t) \cdot i_1(t) + D_1(t) &= E_2(t) \cdot u_2(t-\tau) + G_2(t) \cdot i_2(t-\tau) + H_2(t-\tau) \\ A_2(t) \cdot u_2(t) + B_2(t) \cdot i_2(t) + D_2(t) &= E_1(t) \cdot u_1(t-\tau) + G_1(t) \cdot i_1(t-\tau) + H_1(t-\tau) \end{aligned} \quad (2.12)$$

Where the unknowns at the time $t$ are $u_1(t)$, $i_1(t)$, $u_2(t)$, $i_2(t)$. $A_1(t)$ and $B_1(t)$ are known values. $D_1(t)$ is calculated value and it is the linear combination of the old states of $u_1(s)$ and $i_1(s)$, $s \in [0,t)$. $E_2(t)$ and $G_2(t)$ are known values. $H_2(t-\tau)$ is calculated value and it is the linear combination of the old states of $u_2(s)$ and $i_2(s)$, $s \in [0, t-\tau)$. Similarly, $A_2(t)$ and $B_2(t)$ are known values. $D_2(t)$ is the linear combination of the old states of $u_2(s)$ and $i_2(s)$, $s \in [0,t)$. $E_1(t)$ and $G_1(t)$ are known values. $H_1(t-\tau)$ is the linear combination of the old states of $u_1(s)$ and $i_1(s)$, $s \in [0, t-\tau)$.

(2.12) shows that, the state of $u_1(t)$ and $i_1(t)$ at the time $t$ is depending on the states of $u_2(s), i_2(s)$, $s \in [0, t-\tau]$, and has no relationship with the states of $u_2(s), i_2(s)$, $s \in (t-\tau, t]$. This is straightforward. Because of the propagation delay of the lossy transmission line, the states of $u_2(s), i_2(s)$, $s \in (t-\tau, t]$ are being transferred along the line and have not reached Port 1, so that these states are not able to affect $u_1(t), i_1(t)$ at the time $t$.

There are several simulation techniques exist for the transient analysis of lossy transmission lines. The first kind is the segmentation techniques. In the lumped-RLC method, each segment is represented by as a lumped RLGC network; in the pseudo-lumped method, a lossless transmission line in series with a resistor is used [16][17]. The second kind is the convolution method [18]. The state-based method is an improvement of the convolution method [19][20][21].

## 3. Mimic Transmission Method

The physical circuit is able to work in a distributed manner, and this is mainly because of the existence of transmission lines within the system. First, the transmission line isolates different subcircuits from each other. Second, it has transmission delay. Third, it also helps to stabilize the distributed physical system, since it is passive (lossless or lossy), and it does not bring in any extra energy [25][28].

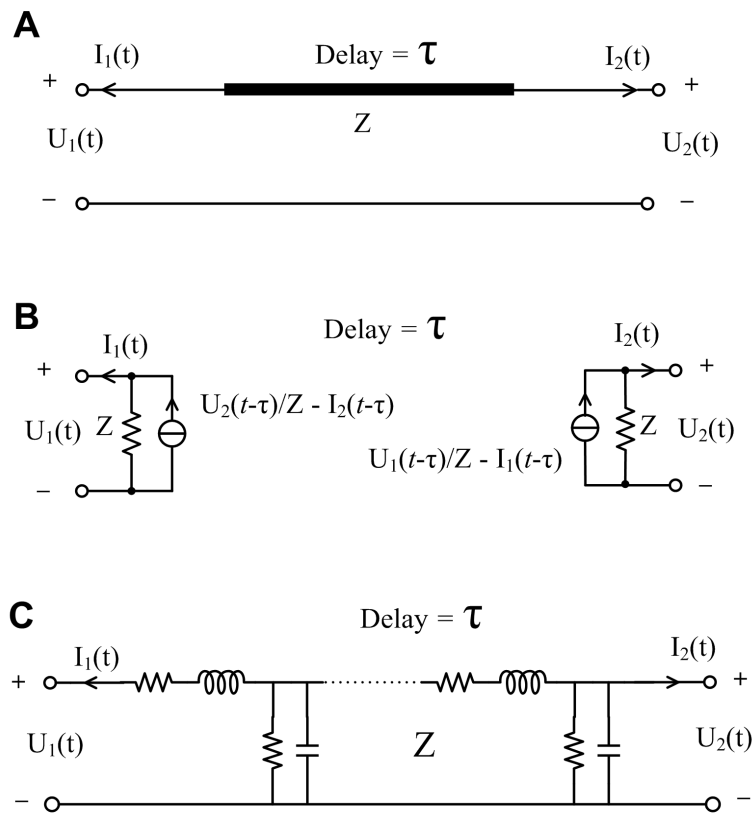

Figure 5. Transmission line. (A) The circuit diagram of the transmission line. (B) The equivalent circuit of the lossless transmission line, according to the Transmission Delay Equations. (C) The equivalent circuit of the lossy transmission line, according

to the lumped-RLGC method.

Our insight into circuit simulation is that we have been aware of the similarity between distributed physical circuit and distributed parallel computer [28]. The transmission delay of wire or interconnect between subcircuits could be mapped to the communication delay of digital data link between processors, as shown in Fig. 6. We suggested emulating the transmission line (lossless or lossy) by the digital data link among processors, because they both have propagation delays. This distributed simulation strategy for transmission line is called Mimic Transmission Method, as shown in Fig. 7.

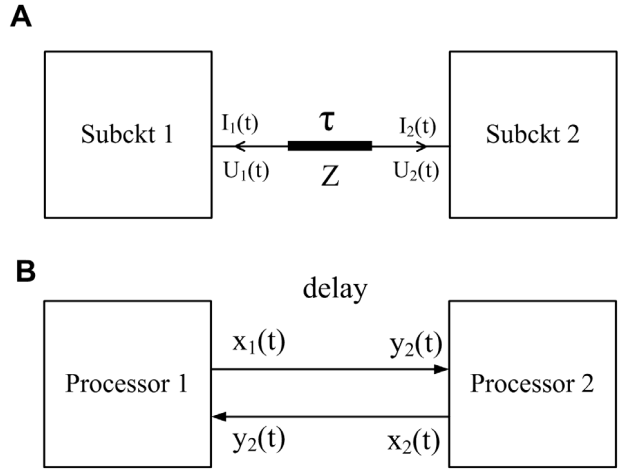

Figure 6. Similarity between physical circuit and parallel computer. (A) The transmission delay between subcircuits. (B) The communication delay between processors.

The mathematic description of the digital data link is the signal delay function:

$$\begin{cases} y_1(t) = x_2(t-T) \\ y_2(t) = x_1(t-T) \end{cases} \quad (3.1)$$

It should be noted that (3.1) is different from (2.8).

To mimic the lossless transmission line, we still need to make some circuit equivalent on the processors. Reformat (2.8) into (3.2) and (3.3):

$$\begin{array}{ll} v_1(t) = u_1(t-\tau) & v_2(t) = u_2(t-\tau) \\ j_1(t) = i_1(t-\tau) & j_2(t) = i_2(t-\tau) \end{array} \quad (3.2)$$

$$\begin{array}{l} u_1(t) + Z \cdot i_1(t) = v_2(t) - Z \cdot j_2(t) \\ u_2(t) + Z \cdot i_2(t) = v_1(t) - Z \cdot j_1(t) \end{array} \quad (3.3)$$

Here (3.2) is the signal delay function and could be emulated by the digital data link. (3.3) could be involved and solved by SPICE on the processor. By this way, the lossless transmission line is emulated by MTM, as shown in Fig. 8.

For the lossy transmission line, the emulation process is similar. First we delay the signals by digital data link:

$$v_1(t) = u_1(t-\tau) \qquad v_2(t) = u_2(t-\tau)$$
$$j_1(t) = i_1(t-\tau) \qquad j_2(t) = i_2(t-\tau)$$

Then (2.12) could be expressed as (3.4). Later we process the delayed signals by SPICE using convolution method or lumped methods on each processor, as shown in Fig. 9.

$$A_1(t) \cdot u_1(t) + B_1(t) \cdot i_1(t) + D_1(t) = E_2(t) \cdot v_2(t) + G_2(t) \cdot j_2(t) + H_2(t)$$
$$A_2(t) \cdot u_2(t) + B_2(t) \cdot i_2(t) + D_2(t) = E_1(t) \cdot v_1(t) + G_1(t) \cdot j_1(t) + H_1(t)$$
(3.4)

The traditional methods, lumped or convolution method, are the technical bases of MTM, but the focus of MTM is different from these traditional methods. The traditional algorithms are used to simulate the transmission lines within SPICE running on a single processor [1][2], while MTM is dedicated to emulate the transmission lines among a number of SPICEs running on a number of processors respectively.

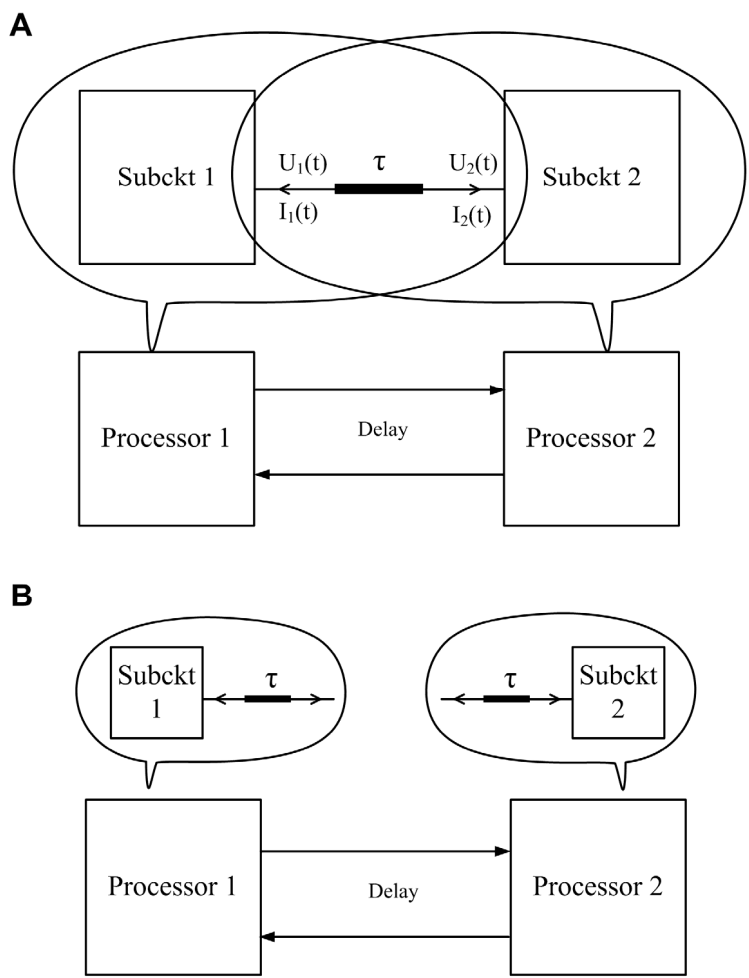

Figure 7. Mimic Transmission Method. (A) Partition the circuit by interfacial wires (or interconnects, transmission lines). (B) Locate each subcircuit onto a processor.

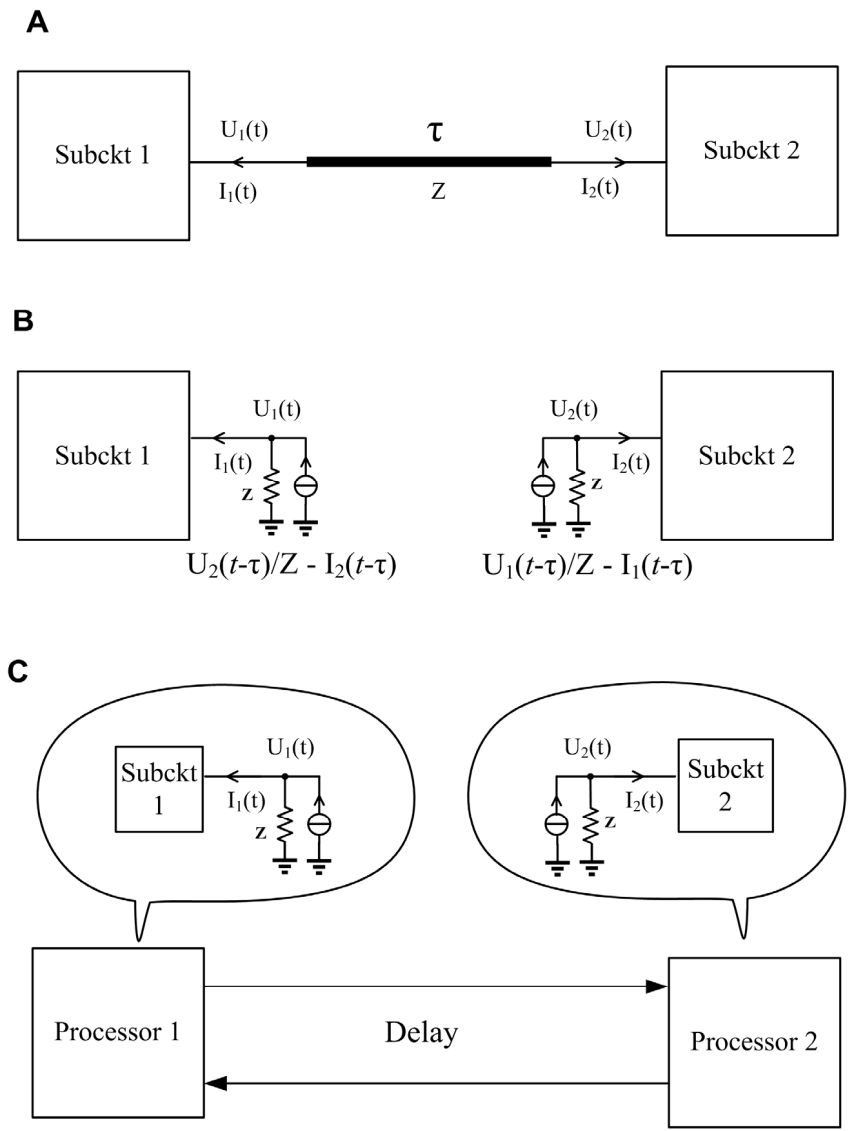

Figure 8. Mimic Transmission Method for lossless transmission line. (A) Physical circuit. One internal lossless transmission line connects two subcircuits. (B) Partition the circuit by lossless transmission line. (C) Locate each subcircuit onto a processor.

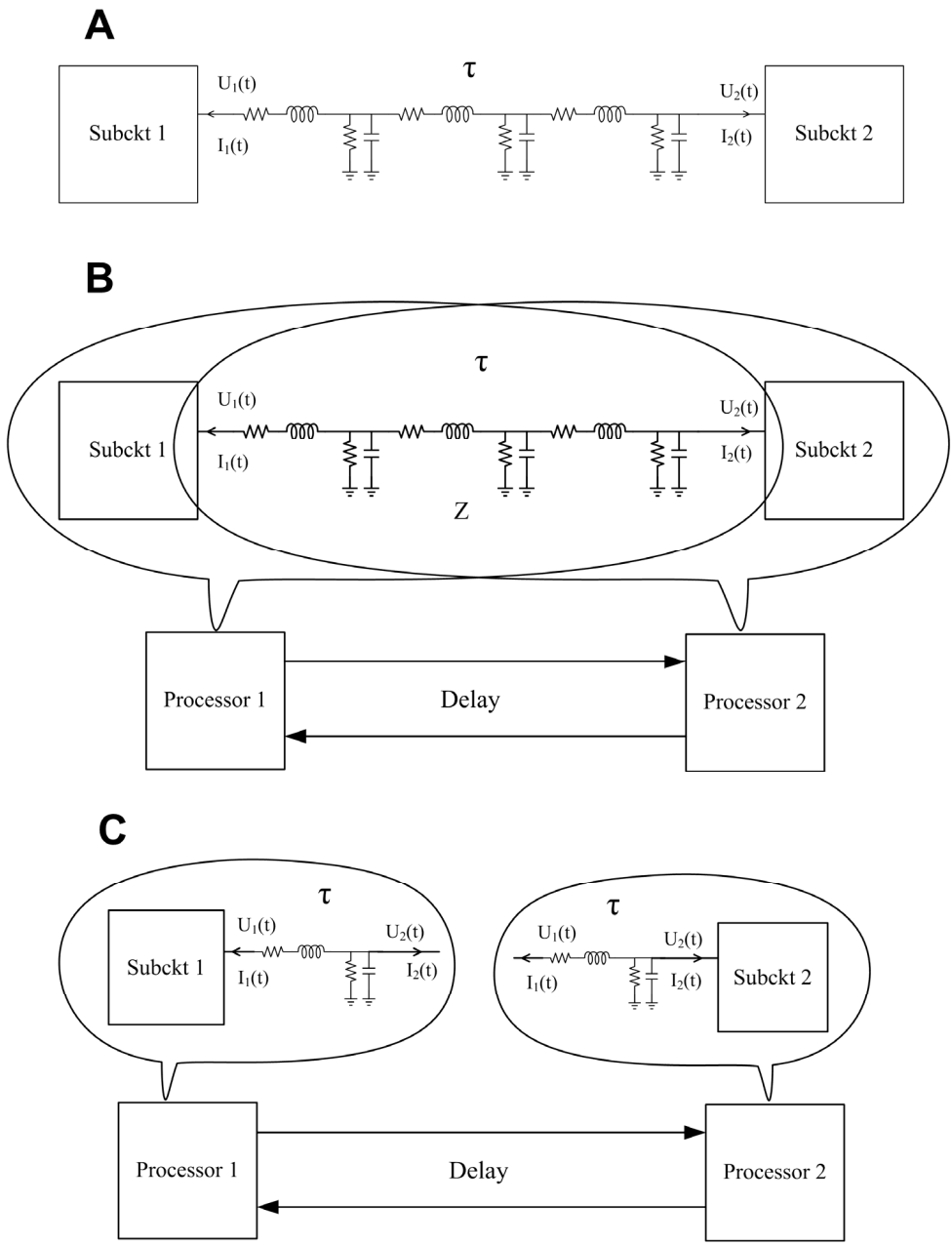

Figure 9. Mimic Transmission Method for lossy transmission line. (A) Physical circuit. An internal lossy transmission line connects two subcircuits. (B) Partition the circuit by lossy transmission line. (C) Locate each subcircuit onto a processor. Note that the interfacial lossy wire is overlapped. Here overlapping is feasible for lossy wire because the analytical description for the lossy wire is linear partial differential equations.

## 4. Example

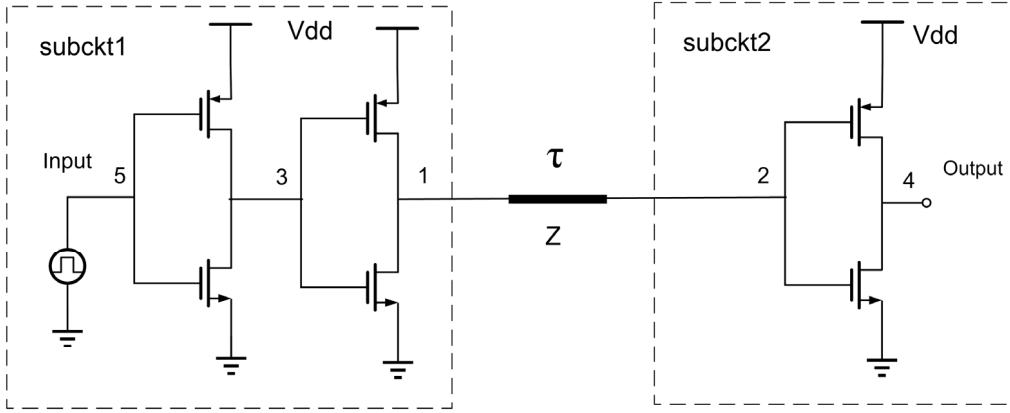

Figure 10. Test circuit. Two subcircuits are connected by a wire. The length of this wire is 1 mm. The frequency of the input digital signal is 1GHz.

The analytical form of Subcircuit 1 is a set of nonlinear ordinary differential equations:

$$f_1(\frac{du_1}{dt}, \frac{du_3}{dt}, \frac{du_5}{dt}, u_1, u_3, u_5, i_1) = 0 \tag{4.1}$$

So is Subcircuit 2:

$$f_2(\frac{du_2}{dt}, \frac{du_4}{dt}, u_2, u_4, i_2) = 0 \tag{4.2}$$

There is a wire connected these two succircuits on the chip, which might be metal or silicide. $L = 4\pi \times 10^{-7} H/m$, $C = \frac{1}{9\pi} \times 10^{-7} F/m$.

Assume that the power Vdd is on at $t = 0^+$, and all the voltages and currents is 0 when $t < 0$.

## 4.1 Lossless wire

If this wire is considered to be lossless, its function is (2.8):

$$u_1(t) + Z \cdot u_1(t) = u_2(t-\tau) - Z \cdot i_2(t-\tau)$$
$$u_2(t) + Z \cdot u_2(t) = u_1(t-\tau) - Z \cdot i_1(t-\tau)$$

Then we use the MTM to solve this circuit on 2 processors. First we set the simulation time window to be $\tau$. The delay function of the digital data link is (3.2).

$$v_1(t) = u_1(t-\tau) \quad\quad v_2(t) = u_2(t-\tau)$$
$$j_1(t) = i_1(t-\tau) \quad\quad j_2(t) = i_2(t-\tau)$$

In the first time window $[0, \tau)$, for Subcircuit 1, we set $v_2(t) = 0, j_2(t) = 0$, because we have the assumption that the $u_2(t) = 0, i_2(t) = 0, t \in [-\tau, 0)$. Then we solve (4.3) to get the waveform of $u_1(t), i_1(t), t \in [0, \tau)$. If we equalize the lossless transmission line into distributed controlled sources, as in Fig. 8B, then (4.3) could be solved by SPICE on Processor 1.

$$\begin{cases} f_1(\dfrac{du_1}{dt}, \dfrac{du_3}{dt}, \dfrac{du_5}{dt}, u_1, u_3, u_5, i_1) = 0 \\ u_1(t) + Z \cdot i_1(t) = v_2(t) - Z \cdot j_2(t) \\ v_2(t) = u_2(t - \tau) = 0 \\ j_2(t) = i_2(t - \tau) = 0, \quad t \in [0, \tau) \end{cases} \quad (4.3)$$

Similary, we get the waveforms of $u_1(t), i_1(t), t \in [0, \tau)$ in Subcircuit 2 by solving (4.4) in SPICE on Processor 2.

$$\begin{cases} f_2(\dfrac{du_2}{dt}, \dfrac{du_4}{dt}, u_2, u_4, i_2) = 0 \\ u_2(t) + Z \cdot i_2(t) = v_1(t) - Z \cdot j_1(t) \\ v_1(t) = u_1(t - \tau) = 0 \\ j_1(t) = i_1(t - \tau) = 0 \quad t \in [0, \tau) \end{cases} \quad (4.4)$$

After this, Processor 1 sends the waveforms of $u_1(t), i_1(t), t \in [0, \tau)$ to Processor 2, and Processor 2 sends the waveforms of $u_2(t), i_2(t), t \in [0, \tau)$ to Processor 1.

$$\begin{array}{ll} v_1(t) = u_1(t - \tau) & v_2(t) = u_2(t - \tau) \\ j_1(t) = i_1(t - \tau) & j_2(t) = i_2(t - \tau) \end{array} \quad t \in [\tau, 2\tau)$$

In the second time window $[\tau, 2\tau)$, for Subcircuit 1, since Processor 1 has received the waveforms of $u_2(t), i_2(t), t \in [0, \tau)$, then $v_2(t), j_2(t), t \in [\tau, 2\tau)$ is known, and we are able to solve (4.5) by SPICE to get the waveforms of $u_1(t), i_1(t), t \in [\tau, 2\tau)$.

$$\begin{cases} f_1(\dfrac{du_1}{dt}, \dfrac{du_3}{dt}, \dfrac{du_5}{dt}, u_1, u_3, u_5, i_1) = 0 \\ u_1(t) + Z \cdot i_1(t) = v_2(t) - Z \cdot j_2(t) \\ v_2(t) = u_2(t - \tau) \\ j_2(t) = i_2(t - \tau) \quad t \in [\tau, 2\tau) \end{cases} \quad (4.5)$$

Similarly, we solve (4.6) by SPICE to get the waveforms of $u_1(t), i_1(t), t \in [\tau, 2\tau]$ in Subcircuit 2.

$$\begin{cases} f_2(\dfrac{du_2}{dt}, \dfrac{du_4}{dt}, u_2, u_4, i_2) = 0 \\ u_2(t) + Z \cdot i_2(t) = v_1(t) - Z \cdot j_1(t) \\ v_1(t) = u_1(t-\tau) \\ j_1(t) = i_1(t-\tau) \qquad t \in [\tau, 2\tau) \end{cases} \qquad (4.6)$$

Repeat this distributed computing process, as illustrated in Fig. 13A. Actually the communication and the computation could be done simultaneously. The result could be sent once it is solved. It should be noted that MTM is not a global iterative algorithm, and we do only one computation at each time window. MTM is a local iterative algorithm because it needs NR iterations to solve the local subcircuit inside SPICE.

We could finally get the simulation result as shown in Fig. 11, which is the same as the simulation result of SPICE on 1 processor.

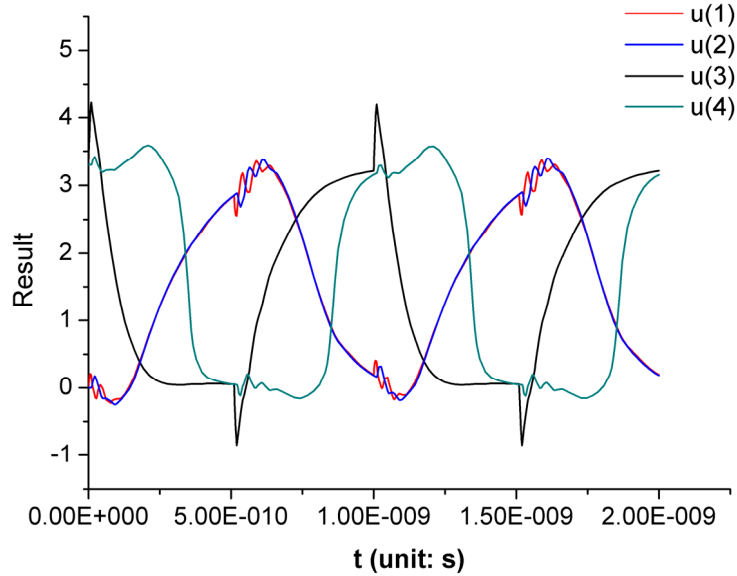

Figure 11. Distributed computing result of MTM on two processors.

### 4.2 Lossy wire

If this wire is lossy, its function is (2.12):

$$A_1(t) \cdot u_1(t) + B_1(t) \cdot i_1(t) + D_1(t) = E_2(t) \cdot u_2(t-\tau) + G_2(t) \cdot i_2(t-\tau) + H_2(t-\tau)$$
$$A_2(t) \cdot u_2(t) + B_2(t) \cdot i_2(t) + D_2(t) = E_1(t) \cdot u_1(t-\tau) + G_1(t) \cdot i_1(t-\tau) + H_1(t-\tau)$$

We use MTM to solve this circuit on 2 processors. First we set the simulation time window to be $\tau$. The delay function of the digital data link is (3.2).

During the first time window $[0, \tau)$, for Subcircuit 1, we set $v_2(t) = 0, j_2(t) = 0$, because we have the assumption that the $u_2(t) = 0, i_2(t) = 0, t \in [-\tau, 0)$. Then we solve

(4.7) to get the waveform of $u_1(t), i_1(t), t \in [0, \tau)$.

$$\begin{cases} f_1(\dfrac{du_1}{dt}, \dfrac{du_3}{dt}, \dfrac{du_5}{dt}, u_1, u_3, u_5, i_1) = 0 \\ A_1(t) \cdot u_1(t) + B_1(t) \cdot i_1(t) + D_1(t) = E_2(t) \cdot v_2(t) + G_2(t) \cdot j_2(t) + H_2(t) \\ v_2(t) = u_2(t-\tau) = 0 \\ j_2(t) = i_2(t-\tau) = 0 \qquad t \in [0, \tau) \end{cases} \qquad (4.7)$$

(4.7) is a theoretical description of the physical circuit with lossy lines. In practice, it is not necessary to calculate the time domain equations of the lossy transmission line (2.12), and we use the lumped-RLGC method or the state-based method to simulate the lossy wire in SPICE, as illustrated in Fig. 9.

Similarly, we solve (4.8) in SPICE and obtain the waveform of $u_1(t), i_1(t), t \in [0, \tau]$ in Subcircuit 2.

$$\begin{cases} f_2(\dfrac{du_2}{dt}, \dfrac{du_4}{dt}, u_2, u_4, i_2) = 0 \\ A_2(t) \cdot u_2(t) + B_2(t) \cdot i_2(t) + D_2(t) = E_1(t) \cdot v_1(t) + G_1(t) \cdot j_1(t) + H_1(t) \\ v_1(t) = u_1(t-\tau) = 0 \\ j_1(t) = i_1(t-\tau) = 0 \qquad t \in [0, \tau) \end{cases} \qquad (4.8)$$

After that, Processor 1 sends $u_1(t), i_1(t), t \in [0, \tau)$ to Processor 2, and Processor 2 sends $u_2(t), i_2(t), t \in [0, \tau)$ to Processor 2.

$$\begin{array}{cc} v_1(t) = u_1(t-\tau) & v_2(t) = u_2(t-\tau) \\ j_1(t) = i_1(t-\tau) & j_2(t) = i_2(t-\tau) \end{array} \qquad t \in [\tau, 2\tau)$$

During the second time window $t \in [\tau, 2\tau)$, for Subcircuit 1, since Processor 1 has received the waveform of $u_2(t), i_2(t), t \in [0, \tau)$, then $v_2(t), j_2(t), t \in [\tau, 2\tau)$ is known, and we are able to solve (4.9) to get the waveform of $u_1(t), i_1(t), t \in [\tau, 2\tau)$.

$$\begin{cases} f_1(\dfrac{du_1}{dt}, \dfrac{du_3}{dt}, \dfrac{du_5}{dt}, u_1, u_3, u_5, i_1) = 0 \\ A_1(t) \cdot u_1(t) + B_1(t) \cdot i_1(t) + D_1(t) = E_2(t) \cdot v_2(t) + G_2(t) \cdot j_2(t) + H_2(t) \\ v_2(t) = u_2(t-\tau) \\ j_2(t) = i_2(t-\tau) \qquad t \in [\tau, 2\tau) \end{cases} \qquad (4.9)$$

Similarly, we solve (4.10) to get the waveform of $u_1(t), i_1(t), t \in [\tau, 2\tau)$ in Subcircuit 2 by SPICE.

$$\begin{cases} f_2(\dfrac{du_2}{dt}, \dfrac{du_4}{dt}, u_2, u_4, i_2) = 0 \\ A_2(t) \cdot u_2(t) + B_2(t) \cdot i_2(t) + D_2(t) = E_1(t) \cdot v_1(t) + G_1(t) \cdot j_1(t) + H_1(t) \\ v_1(t) = u_1(t-\tau) \\ j_1(t) = i_1(t-\tau) \qquad\qquad t \in [\tau, 2\tau] \end{cases} \qquad (4.10)$$

Repeat this distributed computing process and we could finally get the simulation result which is the same as the simulation result of SPICE on one processor.

## 5. Comments

### 5.1 Wire Tearing

Before the distributed simulation of physical circuit by MTM, we need to partition the circuit. First we select some internal wires to be the interfacial wires, and then we tear the circuit into a number of subcircuits connected by these interfacial wires. This is called wire tearing.

The modeling of these interfacial wires is totally based on the parasitic extraction from physical circuits, e.g. layout of integrated circuits.

Sometimes it is not necessary to choose the whole wire, but only part of it to be the interfacial wire. If the transmission delays of all the interfacial wires are same, the synchronization task of MTM would be simple. If the delays of the interfacial wires are different, the synchronization might be complicated.

The advantage of wire tearing over the branch tearing and node tearing is that it does not bring in extra energy since all the wires are passive.

As illustrated in Fig. 12, Branch tearing can be interpreted as the insertion of independent current sources in series with "torn" branches in order to partition the circuit. Node tearing can be interpreted as the insertion of independent voltage sources between "torn" nodes and ground in order to partition the circuit [12][13][14]. So extra energy is brought in by branch tearing and node tearing, which is not natural. This explains why the distributed iterative algorithms based on node tearing or branch tearing are inclined to be unconvergent.

Wire tearing can be interpreted as tearing the circuit by the internal wire inside this circuit (as what we do in this paper), and it could also be interpreted as the insertion of virtual wire in series with "torn" branches in order to partition the circuit (as what we did in VTM, called virtual wire tearing) [25][28].

The efficiency of MTM is depending on the delay of the interfacial wire. The larger the delay is, the longer the simulation time window is. As the result, this distributed algorithm would be more efficient if we select longer internal wires as the interfacial wires. This guideline is straightforward.

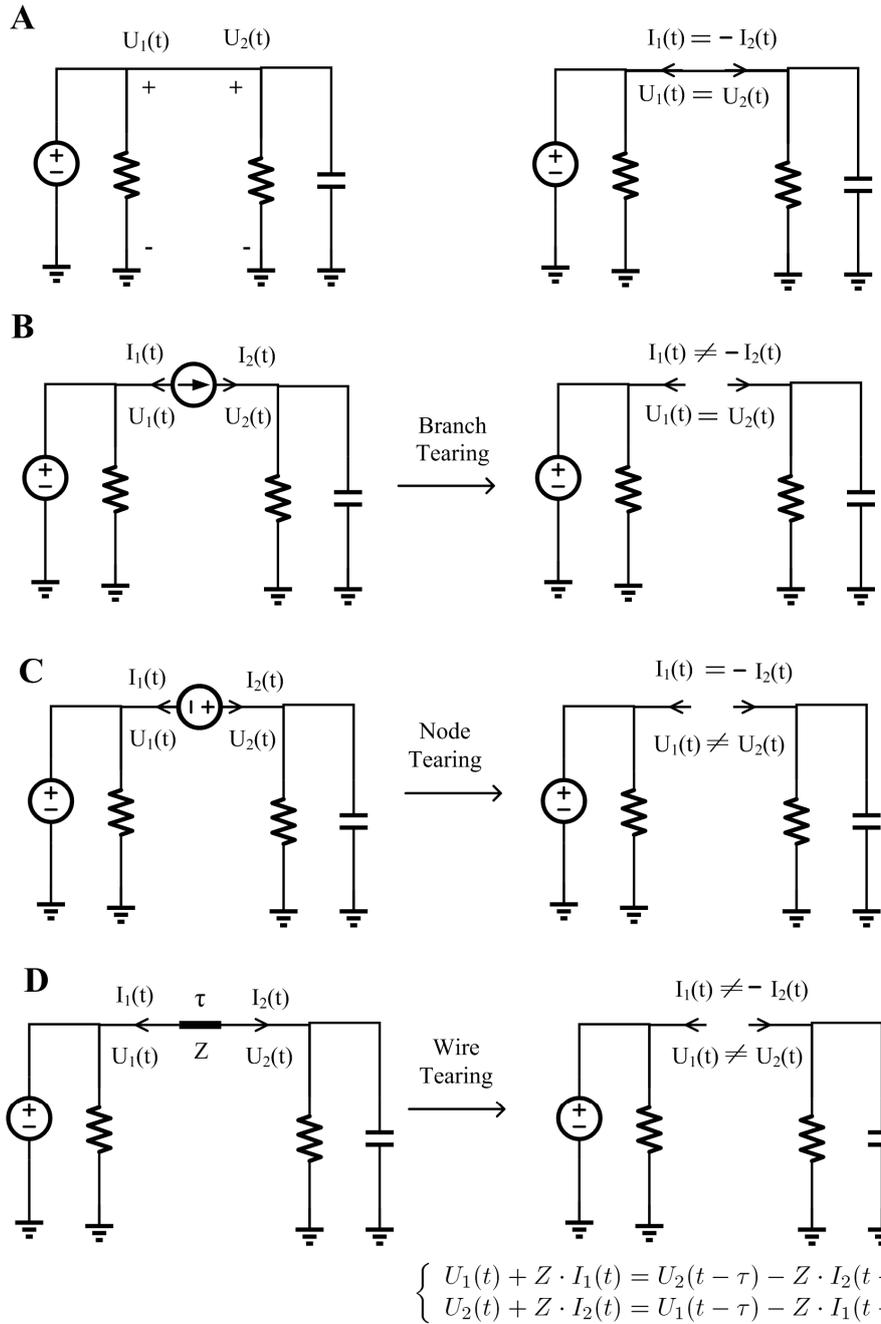

Figure 12. Tearing methods for circiuit. (A) Original circuit. (B) Branch tearing can be interpreted as the insertion of independent current sources in series with "torn" branches in order to partition the circuit. (C) Node tearing can be interpreted as the insertion of independent voltage sources between "torn" nodes and ground in order to partition the circuit. (D) Wire tearing can be interpreted as tearing the circuit by the internal wire inside this circuit, and it could also be interpreted as the insertion of virtual wire in series with "torn" branches in order to partition the circuit.

## 5.2 Simulation Time Step

Theoretically, MTM is able to solve all kinds of physical circuits, as long as the simulation time window of transient analysis is set to be less than the delay of the interfacial interconnects.

$$window \leq \tau$$

In practice, to simulate the system, the time window must be an integral times larger than a single time step of transient analysis:

$$window = K \cdot step, K = 1, 2, 3, \cdots$$

Then, the time window must be larger than a single time step:

$$window \geq step$$

As the result, the delay of interfacial interconnects must be larger than a time step:

$$\tau \geq step$$

Considering the relationship of the length and delay of the transmission line (2.6), we get:

$$\tau = l\sqrt{LC}$$

Here $L$ is the inductance per unit length, and $C$ is the capacitance per unit length.

Finally, we get the following conclusion, to implement MTM, the simulation time step should be set to satisfy:

$$step \leq l\sqrt{LC} \qquad (5.1)$$

## 5.3 The length of interfacial wires

Usually, the time step is determined by the working frequency of the circuit, here $N$ is a constant integer. The value of $N$ is according to the simulation precision.

$$step = \frac{1}{N \cdot f}$$

Therefore,

$$\frac{1}{N \cdot f} \leq l\sqrt{LC}$$

$$l \geq \frac{1}{N \cdot f \sqrt{LC}} \qquad (5.2)$$

Applying (5.2) to certain type of circuit, we might draw some immature conclusions:

For the transistor-level simulation of VLSI with a frequency $f = 1GHz$, if $N$ is set to be 100, then it should be that $l \geq 1mm$. If $l < 1mm$, then $N$ should be larger than 100, which means that the time step is smaller, and the simulation precision is raised.

For the post-layout simulation of analog circuit with a frequency $f = 100MHz$, e.g. ADC or PLL, if $N \approx 1000$, then it should be that $l \geq 1mm$.

To test the signal integrity on PCB board with a frequency $f = 100MHz$, if $N \approx 100$, then it should be that $l \geq 1cm$.

To simulate the power supply network with a frequency $f = 50Hz$, if $N \approx 100$, then it should be that $l \geq 100km$.

Actually, when implementing MTM, the time steps of subcircuits might be different, and the delays of interfacial interconnects might be different. These complicated cases are not considered here.

## 5.4 Performance Modeling

MTM is similar to WR, but essentially they are different. MTM is a global direct algorithm. As shown in Fig. 13, MTM need only one computation at each time window to get accurate result, and the total number of distributed computations is:

$$k_{distri} = \frac{t_2 - t_1}{window} = \frac{t_2 - t_1}{K \cdot step} = \frac{t_2 - t_1}{step} \cdot \frac{1}{K}$$

Usually $K = 1$ for MTM. $k_{distri}$ is also the total number of distributed communications.

For the distributed WR method, assume it needs $k$ waveform iterations at each time window, then the total number of distributed computations is:

$$k_{distri} = \frac{t_2 - t_1}{window} \cdot k = \frac{t_2 - t_1}{K \cdot step} \cdot k = \frac{t_2 - t_1}{step} \cdot \frac{k}{K}$$

Usually $K = 10 \sim 100$ and $k > 10K$.

For the distributed NR method, assume it needs $k$ nonlinear iterations at each time step, and it also needs 2 iterations to solve the nonsymmetrical linear system by Schur complement method, so the total number of distributed computations is

$$k_{distri} = \frac{t_2 - t_1}{step} \cdot 2k$$

Usually $k > 5$.

As the result, MTM would be faster than the distributed NR and WR. NR and WR are iterative algorithms and they might be unconvergent, but MTM is a global direct algorithm and we do not need to worry about the unconvergence problems. This is the main advantage of MTM over the traditional algorithms.

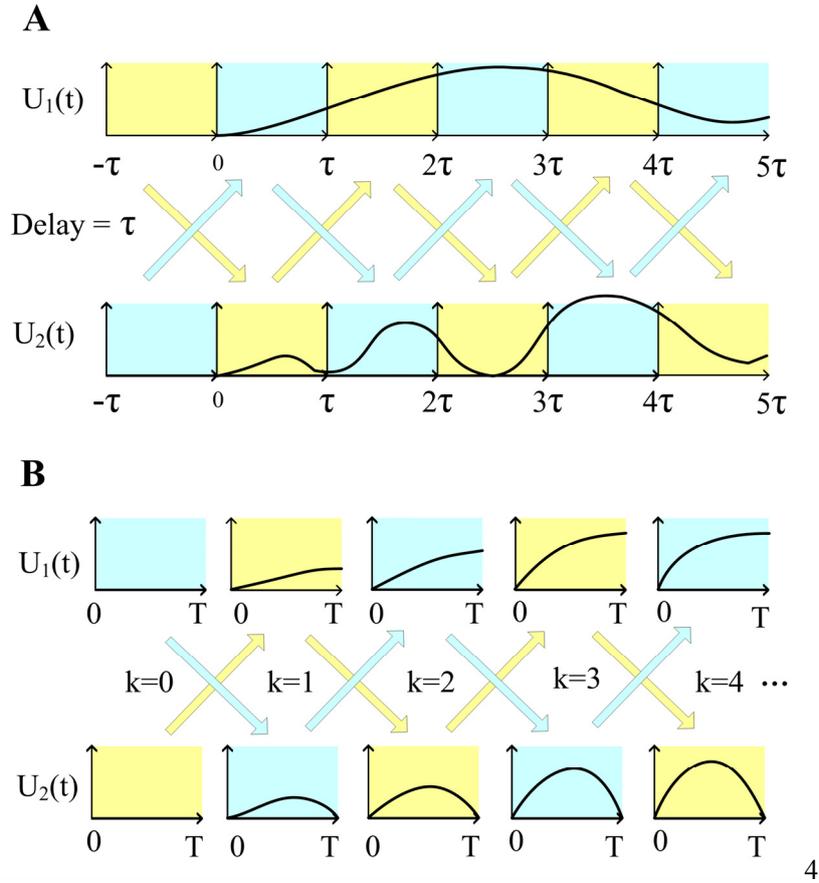

Figure 13. The computing results of MTM and WR. (A) Result of MTM on two processors. MTM is a global direct algorithm and it does only one distributed computation at each time window. (B) Result of WR on two processors. WR is a global iterative algorithm and it does many distributed iterations at one time window until meeting convergence.

## 6. Conclusion

MTM is a simple distributed algorithm mimicking the nature. It totally mimics the distributed behavior of physical circuit by mimicking the distributed behavior of transmission line.

As known, very large integrated circuit (VLSI) was difficult to be simulated in parallel. VLSI produces nonlinear ordinary differential equations and non-symmetrical sparse linear systems, both of which are tough for distributed solvers. Consequently, if we forget the physical background of circuits and treat this problem as pure mathematic equations, then we could only repeat the old road of the pioneers in this research area.

MTM is a black-box algorithm. By mimicking the transmission lines, MTM seals the nonlinear dynamical system within the subcircuit and solve each subcircuit by SPICE, as illustrated in Fig. 3. As the result, we do not need to pay attention on how to solve the nonlinear dynamic system or non-symmetric linear system in parallel. This is different from the traditional work [8][9][11]. MTM makes the distributed circuit simulation straightfoward to be comprehended by electronic engineer. This is the main contribution of our work.

If the modeling of physical system is precise enough, the simulating result of MTM would be accurate, since parasitic elements can make this algorithm more stable and accurate. As the result, MTM is suited for the high-precision transistor-level full-chip post-layout simulation of very large integrated circuits (VLSI), system-on-chip (SoC), network-on-chip (NoC) and system-in-package (SiP). MTM could also be used to simulate the microwave or radio-frequency (RF) circuits and power systems.

The distributed computing strategy from MTM could also be transplanted to solve many other physical problems, e.g. electromagnetics, mechanics or thermodynamics, as long as there are transmission phenomena or wave equations inside these problems.

*Acknowledgement:*

We thank Dr. Qi Wei, Dr. Bo Zhao, Dr. Bin Niu, Dr. Yi Su, Pei Yang, Prof. Yu Wang, Prof. Yongpan Liu, Prof. Fei Qiao, Dr. Xiaojian Mao, Xia Wei, Dr. Xiaozheng Zhong and Prof. Rong Luo.